# Shallow Valence Band of Rutile GeO$_2$ and P-type Doping

Christian A. Niedermeier,[1,*] Keisuke Ide,[1] Takayoshi Katase,[1] Hideo Hosono,[1,2] and Toshio Kamiya[1,2]

*Corresponding author: c-niedermeier@mces.titech.ac.jp

[1]Laboratory for Materials and Structures, Tokyo Institute of Technology, Yokohama 226-8503, Japan

[2]Materials Research Center for Element Strategy, Tokyo Institute of Technology, Yokohama 226-8503, Japan

**Abstract:** GeO$_2$ has an $\alpha$-quartz-type crystal structure with a very wide fundamental band gap of 6.6 eV and is a good insulator. Here we find that the stable rutile-GeO$_2$ polymorph with a 4.6 eV band gap has a surprisingly low ~6.8 eV ionization potential, as predicted from the band alignment using first-principles calculations. Because of the short O–O distances in the rutile structure containing cations of small effective ionic radii such as Ge$^{4+}$, the antibonding interaction between O 2p orbitals raises the valence band maximum energy level to an extent that hole doping appears feasible. Experimentally, we report the flux growth of 1.5 × 1.0 × 0.8 mm$^3$ large rutile GeO$_2$ single crystals and confirm the thermal stability for temperatures up to 1021 ± 10 °C. X-ray fluorescence spectroscopy shows the inclusion of unintentional Mo impurities from the Li$_2$O-MoO$_3$ flux, as well as the solubility of Ga in the r-GeO$_2$ lattice as a prospective acceptor dopant. The resistance of the Ga- and Mo-co-doped r-GeO$_2$ single crystals is very high at room temperature, but it decreases by 2–3 orders of magnitude upon heating to 300 °C, which is attributed to thermally-activated p-type conduction.

## 1. Introduction

To satisfy the worldwide increasing demand for power electronic devices, manufacturers are ramping up the production of more advanced semiconductors than Si with decreased power losses in the high frequency operation range. The enhanced breakdown field strengths and comparably large carrier mobilities of the wide band gap (WBG) compound semiconductors SiC and GaN enable increased energy-efficiency for high frequency and high voltage applications.[1] The commercialization of WBG semiconductors proceeds rapidly, but at the compromise of a significantly higher cost of manufacturing. In particular, large-area SiC and GaN single crystals are very difficult and expensive to grow. Among relevant WBG materials, $Ga_2O_3$ including its β-phase (β-$Ga_2O_3$) is the only oxide semiconductor which is currently being developed for applications in power electronics, because of its high electron mobility and 4.9 eV WBG,[2] with the target of increasing the field strength of devices without triggering electrical breakdown.[3] At the same time, large-area β-$Ga_2O_3$ single crystals are readily available from the standard Czochralski melt growth method,[4] which permits a significant cost advantage. However, many oxide semiconductors including $Ga_2O_3$ are feasible only for n-type doping, thus restricting the development of the current-state WBG oxide semiconductor technology and permitting only the niche application of unipolar devices, such as in Schottky barrier diodes.[5] Therefore, the key challenge is to explore WBG materials for which bipolar doping, both n-type and p-type, can be achieved.

To judge the propensity of carrier doping, it is useful to compare the energetic positions of the band edges of different semiconductors by reference to a common energy scale such as the vacuum level. As a rule of thumb, semiconductors which are easily doped n-type have a large work function (WF) and a conduction band minimum (CBM) that lies shallow in electron energy. Vice versa, those semiconductors which are more easily doped p-type have a low ionization potential (IP) and a valence band maximum (VBM) that lies deep in electron energy.[6] The threshold energy levels for doping are given by the Fermi level pinning energies $E_{pin}^n$ and $E_{pin}^p$ for n-type and p-type doping, which for oxide semiconductors lie at about 3.5 and 6.5 eV below the vacuum level, respectively.[7,8] The Fermi level can hardly be raised above $E_{pin}^n$ or lowered below $E_{\text{pin}}^{\text{p}}$ because intrinsic compensating defects will form spontaneously. The Fermi level pinning energies can be considered roughly constant for materials with a similar chemical affiliation,[9] for example, among the group of oxide semiconductors.[7]

Exclusively all of the established high-mobility oxide semiconductors contain any of the following post-transition metal cations: $Zn^{2+}$, $Cd^{2+}$, $Ga^{3+}$, $In^{3+}$, and $Sn^{4+}$ in the $(n-1)d^{10}ns^0$ electronic configurations, where $n$ is the principal quantum number. Because the energy level of the $ns^0$ orbital-derived CBM lies deep, the WF is typically as large as 4–5 eV,[10] and these oxide semiconductors are very easily doped with electrons as free carriers. In contrast, the reproducible and stable p-type doping and the band conduction by holes as majority carriers at room temperature (RT) have not been achieved. However, through a fundamental understanding of the relationship between the chemical bonding, the structural coordination of atoms and the electronic structure, it is possible to tailor the magnitude of the band gap as well as the energetic positions of the CB and VB of semiconductors. For example, by modifying the crosslinking and the coordination environment of $Cu^+$ cations in the crystal structure and by permitting the strong hybridization between the Cu 3d and O 2p states with similar energy levels, the 3.1 eV WBG p-type semiconductor $CuAlO_2$ has been designed.[11]

Relevant examples of designing semiconductors from insulating oxides are represented by $12CaO \cdot 7Al_2O_3$ and the cubic perovskite $SrGeO_3$. The former enables electron doping by exploiting its particular crystal structure composed of sub-nanometer-sized cages occupied by free $O^{2-}$ anions,[12] and the latter utilizes the deepening of the CBM as a result of the beneficial atomic orbital overlap in a crystal structure of high symmetry.[13] Germanium oxides such as α-quartz-$GeO_2$ (αq-$GeO_2$) and monoclinic $SrGeO_3$ were exclusively known as typical wide-gap insulators built from networks of $GeO_4$ tetrahedra. While monoclinic $SrGeO_3$ is the thermodynamically stable phase at the standard condition, it transforms into the cubic perovskite polymorph with a network of regular $GeO_6$ octahedra at high pressure.[14] Because the high-symmetry octahedral coordination of Ge by O atoms prohibits the hybridization of Ge 4s and O 2p orbitals at the Brillouin zone center,[15–17] the CBM lies comparably deep in energy, converting cubic $SrGeO_3$ into an excellent semiconductor with a 2.7 eV band gap, which is easily doped to high electron concentrations.[13] The single crystal growth of cubic $SrGeO_3$ was recently demonstrated and a remarkably high 3.9 ± 0.5 × $10^2$ cm$^2$/Vs intrinsic RT mobility was derived from the polarization and frequency of optical phonons.[18] However, because the successful synthesis requires 5-GPa high pressure conditions, and the crystalline-to-amorphous transition already occurs at temperatures as low as 100 °C at ambient pressure (AP),[19] the practical application of cubic SrGeO$_3$ is greatly limited. Thus, we target the exploration of AP-stable WBG oxides that can be employed as electronic conductors.

Among the various known germanate compounds, the vast majority of AP-stable structures comprise GeO$_4$ tetrahedra.[20] The germanate structures, in which all of the Ge atoms are densely coordinated by more than 4 O atoms, are very rare. The few exceptions include Zn$_2$Ge$_3$O$_7$[21] and $\beta$-SrGe$_2$O$_5$[22] built from ZnO$_4$-GeO$_6$ and GeO$_5$-GeO$_6$ polyhedra networks respectively, but these metastable structures decompose at temperatures of 600 and 800 °C. In contrast, the rutile polymorph of GeO$_2$ (r-GeO$_2$), a naturally occurring mineral named as argutite,[23] is the only known AP-stable germanate structure with exclusive octahedral coordination of Ge by O atoms while showing a high thermal stability up to 1021 ± 10 °C. At temperatures above, the $\alpha$q-structure built from a network of GeO$_4$ tetrahedra becomes stable, but a prolonged annealing is necessary for complete devitrification (Figure 1).[24] The $\alpha$q- and r-GeO$_2$ structures have the densities of 4.29 and 6.27 gcm$^{-3}$, respectively, and they exhibit a strikingly different chemical stability. $\alpha$q-GeO$_2$ dissolves in acids, bases and hot water, whereas r-GeO$_2$ shows excellent resistance against strong acids like HCl, HNO$_3$, and even HF,[24] demonstrating great suitability for practical applications in semiconductor manufacturing. Besides, r-GeO$_2$ has a remarkably high Vickers hardness number of ~2500,[25] which is larger than that of sapphire, and it is exceeded only by materials which are synthesized at a very high pressure, such as stishovite r-SiO$_2$.[26]

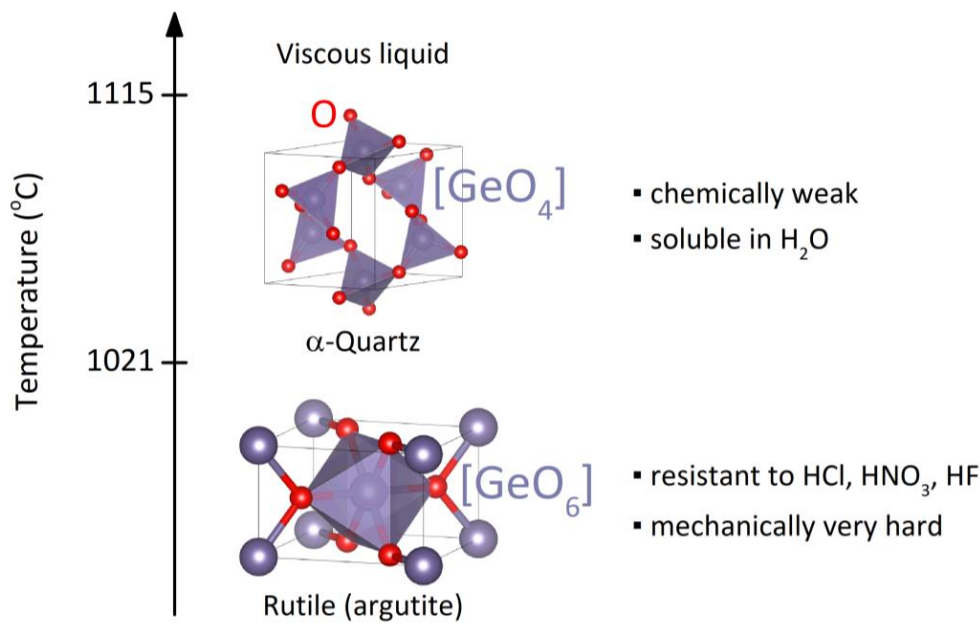

Figure 1: α-Quartz and rutile polymorphs of $GeO_2$. The $\alpha$q-structure forms a network of corner-sharing $GeO_4$ tetrahedra, whereas the rutile structure is built from a network of edge- and corner-sharing $GeO_6$ octahedra, showing excellent chemical stability. The thermodynamically stable phases are r-$GeO_2$ at ambient conditions and $\alpha$q-$GeO_2$ and liquid $GeO_2$ at temperatures above 1021 ± 10 °C and 1115 °C, respectively.

Another beneficial property of $GeO_2$ is the remarkably low melting point (1115 °C) as compared to the established WBG oxide semiconductors ZnO (1975 °C),[27] $\beta$-$Ga_2O_3$ (1820 °C),[28] $In_2O_3$ (1950 °C)[29] and $SnO_2$ (>2100 °C).[30] However, the $GeO_2$ melt is extremely viscous, and upon cooling, it solidifies as an amorphous glass. Hence, crystal growth from the pure $GeO_2$ melt may be impossible. However, the addition of either one of the alkali oxides $Li_2O$ and $Na_2O$ reduces the viscosity, and r-$GeO_2$ single crystals of high purity can be pulled from the $GeO_2$-rich melt by exploiting the low-temperature eutectics in these alkali germanate systems.[31,32] Furthermore, the growth of large-size 3 × 3 × 9 mm³ r-$GeO_2$ crystals has been successfully demonstrated by hydrothermal synthesis.[33,34] The slow-cooling flux growth presents the most facile method as it requires only basic laboratory equipment, but the obtained crystals' dimensions were only about 0.5–1 mm.[35,36] Up till now, the confirmed flux systems for the successful growth of r-$GeO_2$ are limited to the alkali molybdenates $A_2O$–$MoO_3$ (A = Li, Rb, Cs).

In this study, we investigate the electronic structure and the carrier transport in r-GeO$_2$, and target the hole doping because first-principles calculations predict a very shallow VBM of r-GeO$_2$ with an IP of 6.8 eV, which originates from the strong O 2p – O 2p antibonding interaction because of its small O–O distance and the nonbonding nature between O 2p and Ge 4s states. Experimentally, we present the successful LiO$_2$-MoO$_3$ flux growth of 1.5 × 1.0 × 0.8 mm$^3$ large r-GeO$_2$ single crystals and investigate the substitution of Ge by unintentional impurities as well as Ga as an intentional p-type dopant. The electrical properties of r-GeO$_2$ crystals are investigated by photon excitation with ultraviolet (UV) light and temperature-dependent resistance measurements.

## 2. Methods

### 2.1  Computational Details

We performed the theoretical calculations by using the hybrid density functional theory (DFT) and the projector augmented wave method,[37] as implemented in the Vienna Ab initio Simulation Package (VASP).[38,39] All calculations were carried out by using the experimentally determined lattice parameters, a plane wave cutoff energy of 500 eV and a Γ-centered *k*-mesh with a maximum grid spacing of 0.3 Å$^{-1}$. We used the Perdew-Burke-Ernzerhof (PBE0) hybrid functional[40] for the calculations of r-SnO$_2$, r-GeO$_2$, and αq-GeO$_2$ and employed maximally localized Wannier functions[41] to analyze the electronic band structures. For r-TiO$_2$, we used the Heyd-Scuseria-Ernzerhof (HSE06) hybrid functional[42] with the standard mixing parameter of 25% for the exact-exchange term. The respective hybrid functionals were chosen to obtain calculated band gaps in good agreement with the experiment (Table 1).

The bond-weighted electron density of states (DOS) was calculated by partitioning the band structure energy into bonding, nonbonding, and antibonding contributions and obtaining the crystal orbital Hamilton population (COHP) spectrum.[43,44]

Table 1: Calculated direct band gap energies $E_g$ of r-SnO$_2$, r-TiO$_2$, r-GeO$_2$, and αq-GeO$_2$ by using the PBE0 and HSE06 hybrid functionals in comparison with the experimental results obtained at the respective temperature $T$.

| Material | Direct band gap energy $E_g$ (eV) | | | Temperature $T$ (K) |
|---|---|---|---|---|
| | PBE0 | HSE06 | Experiment | |
| r-SnO$_2$ | 3.8 | 3.1 | 3.56[a,⊥] | 4.5 |
| r-TiO$_2$ | 4.2 | 3.4 | 3.03[b,⊥] | 1.6 |
| r-GeO$_2$ | 4.6 | 3.1 | 4.68[c,⊥] | 10 |
| αq-GeO$_2$ | 6.4 | 5.7 | 6.6[d,⊥] | 298 |

⊥Determined for light with a polarization vector $\vec{E}$ perpendicular to the crystallographic $c$-axis $\vec{c}$ of both, the rutile and α-quartz structures.
[a]Reference[45]
[b]Reference[46]
[c]Reference[47]
[d]Reference[48]

## 2.2 Experimental Details

R-GeO$_2$ single crystals were grown using the slow-cooling flux method. αq-GeO$_2$ powder (0.35 g, 99.99 % purity) was mixed with Li$_2$CO$_3$ (2.63 g, 99.99 % purity) and MoO$_3$ (7.69 g, 99.98 % purity) as a flux, corresponding to a Li$_2$O · 1.5 MoO$_3$ molar composition and a 1:25 weight ratio of GeO$_2$ to flux after the Li carbonate decomposition. The eutectic temperature in the Li$_2$MoO$_4$–Li$_4$Mo$_5$O$_{17}$ system lies at 525 °C.[49] Ga$_2$O$_3$ powder (16.5 mg) was added to investigate the substitutional doping, corresponding to a 5 at. % concentration of Ga with respect to Ge atoms. The mixed powders were pressed to a pellet and melted in a Pt crucible, which is contained within a covered alumina crucible for improved thermal isolation upon slow cooling. After the homogenization of the melt at 980 °C for 1 h, the temperature was slowly decreased at the rate of 3.5 °C/h to 600 °C, from which point the set temperature was ramped down quickly. The flux was washed out with hot deionized water and r-GeO$_2$ single crystals were obtained exclusively. The slow cooling from elevated temperatures above 980 °C was unsuitable, because an increasingly large amount of αq-GeO$_2$ single crystals was obtained.

The out-of-plane high-resolution X-ray diffraction (HR-XRD) pattern of the r-GeO$_2$ single crystal (110) surface was recorded using a Rigaku SmartLab diffractometer equipped with a monochromated Cu K$_{\alpha 1}$ X-ray source (1.5406 Å) and by utilizing parallel beam optics. The nonpolarized Raman spectra were measured in the backscattering configuration using a HORIBA LabRAM HR system and employing the 457 nm wavelength of a Nd:YVO$_4$ laser (Cobolt Twist 50) and a cooled charge-coupled device detector. The thermal stability of single crystals was investigated by annealing in air on a Pt foil and calibrating the furnace temperature with reference to the melting of a small piece of Au (1064 °C). Spectroscopic ellipsometry measurements (HORIBA Jobin Yvon UVISEL) from 0.6 to 6 eV were taken at a 65° incident angle and by using the naturally grown (110) facet as the reflecting surface. An alignment of the [001] optical axis of r-GeO$_2$ with respect to the plane of incidence was not performed because of the difficulty arising from the small crystal size. Wavelength-dispersive X-ray fluorescence (XRF, Bruker S8 Tiger 1 kW) spectra were recorded using a Rh X-ray source and a LiF (200) analyzer crystal. The quantitative XRF analysis of r-GeO$_2$ single crystals was performed with reference to a mixed-phase r-GeO$_2$/Ga$_2$O$_3$/MoO$_3$ polycrystalline pellet sintered at 500 °C for 12 h, with a 94.9, 2.6 and 2.5 at.% composition of Ge, Ga and Mo, respectively, to take into account the enhanced fluorescence of Ga atoms due to the excitation with characteristic Ge K$_\alpha$ X-rays from the r-GeO$_2$ matrix.

The r-GeO$_2$ single crystals were embedded in acrylate resin on glass substrates, polished, and then patterned by the standard photolithography method. 10 nm Ti/50 nm Au contacts were deposited by electron beam evaporation to form 15 × 200 and 50 × 500 µm$^2$ channels for photocurrent and temperature-dependent two-point-probe resistance measurements, respectively. The electrical measurements were carried out using a semiconductor parameter analyzer (Agilent 4155C) and a non-monochromated ultra-high-pressure Hg lamp light source (USHIO SX-UI 501HQ, 67 mW/cm$^2$ intensity) for photon excitation with the 254, 302, 313 and 365 nm emission lines representing the most energetic transitions in the UV spectral region. The temperature-dependent resistance was measured in the dark by probing the single crystal upon heating from RT to 300 °C in air.

## 3. Results and Discussion

### 3.1 Electronic Structure and Band Alignment

R-GeO$_2$ has a tetragonal unit cell and belongs to the space group *P*4$_2$/*mnm* (no. 136). The structure is built from a dense network of slightly distorted, edge- and corner-sharing GeO$_6$ octahedra with O atoms in trigonal-planar coordination with Ge atoms (Figure 2a). The PBE0-calculated direct band gap is 4.6 eV (Figure 2b), which is in agreement with an optical absorption measurement (4.68 eV at 10 K), and it is directly-allowed only for light with a polarization vector $\vec{E}$ perpendicular to the crystallographic *c*-axis $\vec{c}$.[47] The symmetry analysis yields a 5.4 eV dipole-allowed transition from the band at about 0.8 eV below the VBM for $\vec{E}$ parallel to $\vec{c}$.[50] Using the band curvatures, the calculated effective masses for electrons and holes parallel to $\vec{c}$ are 0.23 and 1.6 $m_0$, respectively, where $m_0$ denotes the electron rest mass.

The CBM is mainly derived from Ge 4s states, while the VBM exclusively comprises O 2p states without the contribution of Ge states. The hybridization is different from the established oxide semiconductors like ZnO, Ga$_2$O$_3$, In$_2$O$_3$, and SnO$_2$, for which the unoccupied metal cations' s states partially contribute to the VBM, thus deepening its energy level because of the bonding interaction with O 2p states. Considering the threefold coordination of O to Ge atoms in direct analogy to r-TiO$_2$, the hybridized O bonding orbitals have sp$^2$-character.[51] By visualizing the partial charge density of the top energy band of the VB with the yellow surfaces shown in Figure 2a, the O 2p lone pair orbitals with the lobes oriented parallel along the [110] and [1$\bar{1}$0] crystallographic directions are depicted. These lone pairs may be denoted as O 2p$_{z'}$ orbitals, where z' is the local coordinate, indicating the two directions perpendicular to the (110) and (1$\bar{1}$0) planes containing the threefold coordinated O atoms in the planar OGe$_3$ structural units. By referring to the O 2p$_{z'}$ orbitals as *lone pairs*, we like to point out that these do not contribute to the bonding to Ge atoms, although using this terminology shall not exclude the interaction between the O 2p$_{z'}$ orbitals themselves. Because of the octahedral distortion in the rutile structure, two types of dissimilar O–O interatomic distances denoted as $d_1$ and $d_2$ and two types of similar M-O bond lengths are considered. These distances in r-SnO$_2$, r-TiO$_2$, and r-GeO$_2$ noticeably decrease in this sequence of decreasing effective cation radii (Figure 2c).[52] Surprisingly, in r-GeO$_2$, the O–O distances $d_1$ = 2.41 and $d_2$ = 2.67 are shorter than twice the effective O$^{2-}$ anion radius of 2.72 Å[52] for oxides with a prevalently ionic character of chemical bonds. Because the VBM is exclusively characterized by the interaction between O 2p orbitals, the shortening of these O–O distances will have a major impact on its energy level.

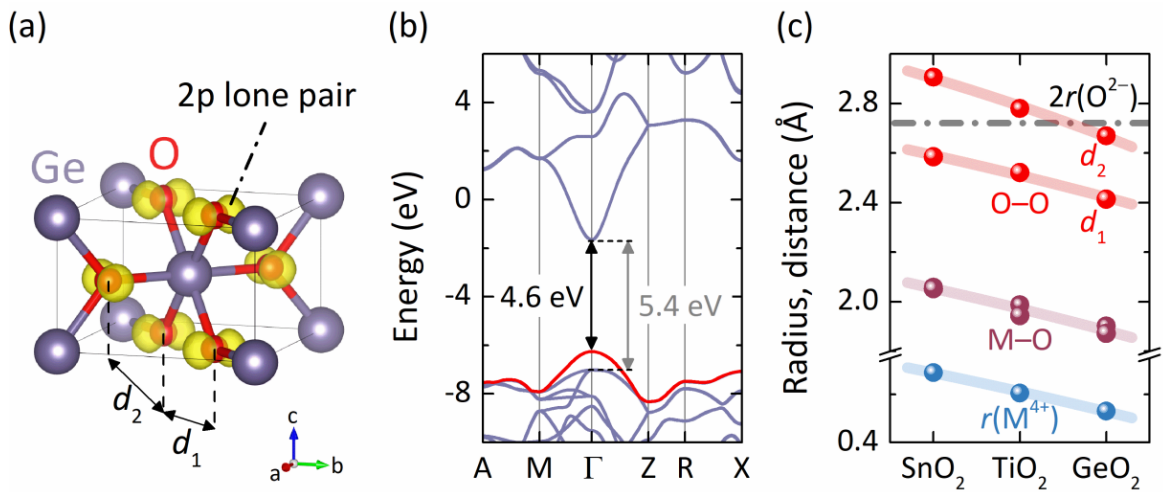

Figure 2: (a) r-GeO$_2$ crystal structure with O atoms in threefold planar-trigonal coordination, forming sp$^2$-hybridized bonds with Ge atoms. The yellow surfaces depict the partial charge density of O 2p lone pair orbitals perpendicular to the OGe$_3$ coordination plane, forming the top VB (marked in red). (b) PBE0-calculated band structure of r-GeO$_2$. (c) Shortest and second-shortest O–O interatomic distances, denoted as $d_1$ and $d_2$, respectively, in the rutile structures become substantially shorter with a decreasing effective M$^{4+}$ cation radius in the sequence Sn$^{4+}$ – Ti$^{4+}$ – Ge$^{4+}$.

We calculated the band structures of r-SnO$_2$, r-TiO$_2$, and r-GeO$_2$ to compare the electronic properties. The energy scale is referenced to the vacuum level $E_{vac}$ by adopting the experimentally determined 8.5 eV IP of SnO$_2$,[53] which is in agreement with the other reported values of 8.3 eV,[10] as well as the range of 7.9–8.9 eV determined for SnO$_2$ thin films,[54] depending on the deposition conditions. In contrast to the covalent bonds between O$^{2-}$ anions and the tetrahedrally coordinated Ge$^{4+}$ cations in αq-GeO$_2$, the bonds in the rutile polymorph with the octahedral cation coordination show a stronger ionic character. Thus, the deep O 2s core level in the partial DOS is used to align the band structures of all rutile phases to a common energy level and determine the r-TiO$_2$ and r-GeO$_2$ IPs (Figure 3a,b). With the decreasing O–O interatomic distances in the rutile structures, the band alignment shows a pronounced shallowing of the VBM from −8.5 eV for r-SnO$_2$ to −6.8 eV for r-GeO$_2$.

Exclusively O 2p states contribute to the DOS at the VB top, and the chemical bonding between O atoms separated by the shortest and second-shortest interatomic distances, $d_1$ and $d_2$, respectively, is analyzed by the COHP spectra (Figure 3c). In the region of the VBM top, the interaction between O atoms separated by $d_1$ is nonbonding (COHP is about zero), whereas the interaction between O atoms separated by $d_2$ is antibonding (–COHP is negative), and the latter one is noticeably stronger in r-GeO$_2$ as compared to that in r-SnO$_2$. Moreover, the orbital-resolved partial COHP indicates a negligible interaction between the O 2p lone pairs oriented *parallel* to each other. The analysis further indicates a pronounced antibonding interaction between the two O 2p lone pairs with the lobes oriented along the [110] and [1$\bar{1}$0] directions *perpendicular* to each other, which are stacked by the distance $c/2$ parallel along the crystallographic $c$-axis (cf. Figure 2a). Thus, the trend in the shallowing of the VBM from r-SnO$_2$ to r-GeO$_2$ is attributed to the decrease in the *second* shortest O–O distance $d_2$, resulting in a strong antibonding interaction between O 2p orbitals with the lobes oriented *perpendicular* to each other.

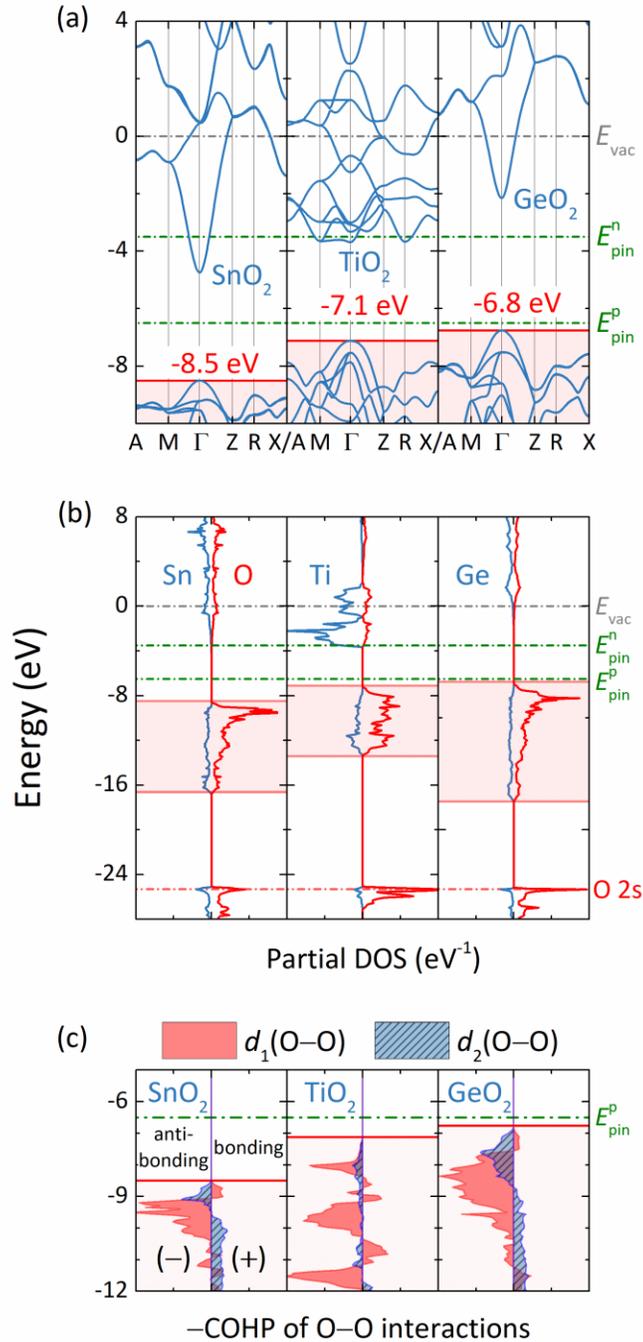

Figure 3: (a) Electronic band structures of r-$SnO_2$, r-$TiO_2$ and r-$GeO_2$. The energy scale is referenced to the vacuum level $E_{\text{vac}} = 0$. (b) Corresponding partial DOS is used for the band structure alignment with reference to the deep O 2s core level. The VBM increases from −8.5 eV for $SnO_2$ to −6.8 eV for $GeO_2$, indicating the stronger propensity for p-type doping. (c) −COHP spectra at the VB top indicate that the interaction between O atoms with the interatomic distance $d_1$ ($d_2$) is nonbonding (antibonding) at the VBM, where $d_1$ and $d_2$ are defined as depicted in Figure 2a.

The −8.5 eV deep VBM for r-SnO$_2$ suggests excellent n-type semiconducting properties, whereas p-type doping is extremely difficult to achieve. The r-TiO$_2$ VBM at −7.1 eV lies noticeably higher in energy, and while the p-type doping of r-TiO$_2$ has been reported,[55,56] the hole conduction in semiconducting oxides without the hybridization of the localized O 2p states with cation states such as Cu 3d in Cu$_2$O[57] or Sn 5s in SnO[58] is controversially discussed in the literature and often reviewed critically. The ~6.8 eV IP of r-GeO$_2$ is yet remarkably low as compared to the established WBG oxide semiconductors ZnO, Ga$_2$O$_3$, In$_2$O$_3$ and r-SnO$_2$. For example, considering the 4.0 eV WF[59] and the 4.9 eV band gap of β-Ga$_2$O$_3$,[2] the estimated IP is ~8.9 eV, indicating that p-type doping is extremely difficult. In contrast, the r-GeO$_2$ IP is comparable to that of GaN (6.7 eV),[60] a WBG semiconductor which at first seemed difficult to dope p-type, but the activation of holes as charge carriers originating from Mg acceptors has first been realized by low-energy electron beam treatment[61] and more practically, by thermal annealing in a N$_2$ ambient to remove H from H–Mg complexes.[62] Moreover, the thin film growth would permit to adjust the deposition parameters for tuning of the IP within an energy window of about 1 eV, as has been demonstrated for r-SnO$_2$ (7.9–8.9 eV).[54]

Regarding the n-type conduction in r-GeO$_2$, the large CB dispersion and the low effective mass of about 0.23 $m_0$ are beneficial for the high mobility of electrons, which are unlikely to be trapped as small polarons at donor atoms. However, the CBM lies at ~1.4 eV above the Fermi energy pinning level $E_{\text{pin}}^{\text{n}}$, indicating that n-type doping is very difficult. This result again illustrates the dilemma of doping >4 eV WBG semiconductors, for which typically either n-type or p-type conduction can be realized. Nevertheless, it is worthwhile to investigate the doping of r-GeO$_2$ experimentally because the band edge position relative to the Fermi level pinning energy merely provides a guideline to judge the doping propensity.

## 3.2 Crystallographic Characterization and Thermal Stability

r-GeO$_2$ single crystals with a maximum size of 1.5 × 1.0 × 0.8 mm$^3$ (Figure 4a) were grown by precipitation from a Li$_2$O·1.5 MoO$_3$ flux, and by only using a fraction of the flux charges reported previously.[35,36] The crystal facets with the largest surface area were identified as the (110) crystallographic planes by HR-XRD out-of-plane measurements (Figure 4b). After the rotational alignment to observe the 221 in-plane diffraction spot (Figure S1), it is found that the fourfold rotational symmetry axis along the square rod-shaped crystals corresponds to the crystallographic *c*-axis of r-GeO$_2$. The results indicate that at the employed growth conditions, the surface energy may be the largest for the (001) planes and the smallest for the (110) planes. In agreement with an analysis of micro-crystalline r-GeO$_2$,[63] the 169, 700, and 870 cm$^{-1}$ phonon peaks are observed in the nonpolarized Raman spectrum measured perpendicular to the crystallographic *c*-axis (Figure 4c). The r-GeO$_2$ single crystals are thermally stable during annealing for 1 h at temperatures up to 1021 ± 10 °C in air. At elevated temperatures, the formation of micrometer-sized droplets on the crystal surface indicates the onset of partial melting. The broad 431 cm$^{-1}$ Raman peak measured for a surface droplet formed after annealing at 1041 °C for 1 h is assigned to the symmetric stretching of bridging O atoms in six-membered GeO$_4$ rings[64] and confirms the formation of amorphous GeO$_2$ (Figure 4d). The characteristic Raman peaks are absent in *α*q-GeO$_2$ as confirmed by comparison with the spectrum of a single crystal (Figure 4e). After annealing at 1066 °C for 12 h, the melting of r-GeO$_2$ is still incomplete, while the characteristic Raman peaks of the crystalline *α*q-phase are not observed, indicating that the rutile-to-*α*-quartz phase transition is kinetically hindered.

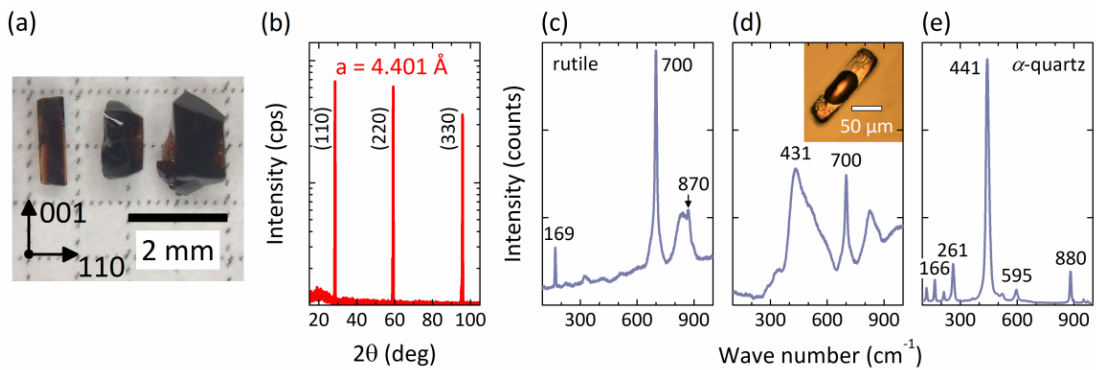

Figure 4: (a) Optical micrograph of r-GeO$_2$ single crystals grown by the Li$_2$O-MoO$_3$ flux method. (b) Out-of-plane HR-XRD pattern of the single crystal (110) surface with the intensity plotted on a logarithmic scale. The Raman scattering spectra in the back-scattering geometry are measured (c) perpendicular to the crystallographic *c*-axis of an as-grown r-GeO$_2$ single crystal, (d) for a surface droplet formed on a r-GeO$_2$ single crystal after annealing at 1041 °C for 1 h in air, and (e) for an *α*q-GeO$_2$ single crystal.

### 3.3 Optical Properties and Crystal Purity

The r-GeO$_2$ absorption coefficient $\alpha$ was measured by spectroscopic ellipsometry, using the (110) crystal facet as the reflecting surface. From the linear fit to the $\alpha^2$–photon energy plot in the vicinity of the absorption edge, a 4.6 eV direct optical band gap is obtained (Figure 5a). Moreover, a noticeable subgap absorption is observed above 3.5 eV and the brown crystal color indicates the formation of defects or the inclusion of impurities. Because the brown color persists after annealing the crystals at 1000 °C for 2 h in air, the formation of intrinsic defects such as O vacancies or Ge$^{2+}$ cations with a reduced oxidation state is unlikely. Instead, the XRF analysis confirms the extrinsic impurity Mo (Figure 5b). Because the brown color is also observed for crystals grown without the addition of Ga$_2$O$_3$ as a dopant, it is attributed to the Mo impurities. Because the effective ionic radius of Mo$^{6+}$ (0.59 Å) is very similar to that of Ge$^{4+}$ (0.53 Å),[52] incorporation into the r-GeO$_2$ lattice during the flux growth cannot be avoided. However, most importantly, the intentional impurity doping of Ga$^{3+}$ (0.62 Å) cations into the r-GeO$_2$ lattice as a prospective p-type acceptor defect is confirmed. From the background-corrected and integrated K$_\alpha$ peak areas, we determine the Ga, Ge and Mo concentrations of 2.6, 96.9 and 0.5 at. %, respectively. Moreover, Al$^{3+}$ (0.54 Å) doping of r-GeO$_2$ was attempted, but the addition of even a small amount of Al$_2$O$_3$ powder to the charge resulted in the precipitation of exclusively $\alpha$q-GeO$_2$ single crystals at the investigated growth conditions.

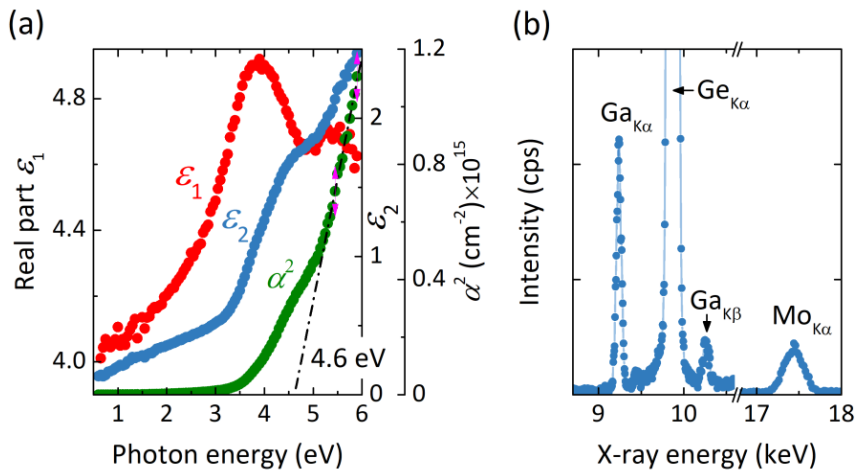

Figure 5: (a) Real $\varepsilon_1$ and imaginary $\varepsilon_2$ parts of the r-GeO$_2$ dielectric function measured by spectroscopic ellipsometry, using the (110) crystal facet as the reflecting surface. The square of the absorption coefficient data $\alpha^2$ indicates a 4.6 eV direct optical band gap and the presence of sub-gap defect states above 3.5 eV. (b) XRF spectrum analysis indicates the inclusion of 2.6 at.% Ga and 0.5 at.% Mo impurities into the r-GeO$_2$ lattice.

## 3.4 Electrical Properties

To investigate the electrical properties of r-GeO$_2$, the single crystals containing unintentional Mo impurities were patterned by the standard photolithography method with two Au/Ti contacts forming a narrow channel (Figure 6a). Because the as-grown single crystals are very resistive, the current response is measured after excitation with a non-monochromated, high-intensity, and ultra-high-pressure Hg lamp as the UV light source. The 4.3 eV WF[65] of the Ti contact allows us to measure the hole conduction, and a good ohmic behavior is observed, while the electron injection would be blocked because of the large offset with the CBM. A clear photocurrent is observed upon switching the UV light source on and off in 10 s time intervals, although the current is as small as ~0.1 nA at a 40 V bias (Figure 6b). Using a 300 nm wavelength filter, which is transparent for light with photon energies of 4.13 and < 2.3 eV, the photocurrent weakens but it persists to a level of ~40 % in accordance with the decreased light intensity. This result indicates that the comparably small photocurrent cannot be attributed to the 4.6 or 5.4 eV band-to-band excitations of carriers.

Lastly, the temperature-dependent resistance of the Mo-doped GeO$_2$ single crystals with and without intentional Ga impurities was measured in the dark. The resistance decreases by about 2–3 orders of magnitude from 4×10$^{12}$ to 5×10$^9$ Ω after heating from RT to 300 °C in air. From the Boltzmann constant $k_B$ and the linear slope $E_a/k_B$ in the range of 125–300 °C in the Arrhenius plot for the resistance, we estimate the thermal activation energies $E_a$ of 0.17 and 0.18 eV, for Mo-doped and Ga- and Mo-codoped r-GeO$_2$ single crystals, respectively (Figure 6c). Considering the large subgap absorption observed in the optical spectrum (cf. Figure 5a), the increase in conductivity upon heating may likely result from a Mo impurity band within the band gap. Because the investigation of Ga as a suitable acceptor defect necessitates the elimination of the unintentional Mo impurities in r-GeO$_2$ crystals, the search for a suitable flux containing only cations with an effective ionic radius dissimilar to that of Ge$^{4+}$ is the subject of ongoing work.

When r-GeO$_2$ single crystals of high purity are grown, it would be exciting to investigate the acceptor doping of r-GeO$_2$ by employing common strategies for WBG semiconductors, for example, the incorporation of H during the growth with subsequent post-treatment by low-energy deionizing electron beam irradiation or thermal annealing, which supports the activation of carriers trapped at in-gap defects.[61,62] A computational study of r-GeO$_2$ further suggests that H forms complexes with Al and Ga acceptor defects, which reduces their formation energy.[66]

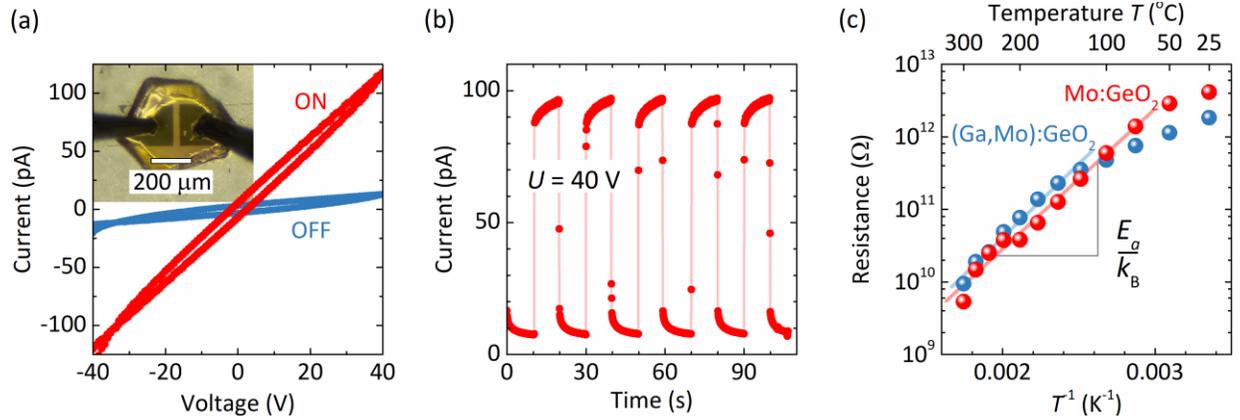

Figure 6: (a) Current-voltage testing of a r-GeO$_2$ single crystal with unintentional Mo impurities under the dark condition (OFF) and upon exposure to a non-monochromated UV light source (ON). The inset shows the 2-point-probe configuration with patterned Au/Ti contacts. (b) Time-dependent photocurrent measurement at a 40 V bias recorded by switching the UV light source on and off in 10 s time intervals. (c) Electrical resistance of Mo-doped and Ga- and Mo-codoped r-GeO$_2$ single crystals decreases by about 2–3 orders of magnitude upon heating from RT to 300 °C.

## 4. Conclusions

The alignment of the r-GeO$_2$ electronic band structure with those of r-SnO$_2$ and r-TiO$_2$ indicates that the doping will be challenging. However, the unusually low 6.8 eV IP suggests that p-type conduction is more realistic to achieve for r-GeO$_2$ as compared to other established WBG oxide semiconductors. The shallow VBM is a result of the strong O 2p antibonding interaction between the lone pair orbitals oriented perpendicular to each other, which is attributed to the particular threefold coordination of O with the Ge atoms and the short O–O distances in the rutile structure with a comparably small effective Ge$^{4+}$ cation radius.

We report the flux growth of large-size, 1.5 × 1.0 × 0.8 mm$^3$ r-GeO$_2$ single crystals and confirm the thermal stability for temperatures up to 1021 ± 10 °C. The XRF analysis confirms the inclusion of unintentional Mo impurities from the flux and the solubility of Ga atoms in the r-GeO$_2$ lattice as prospective acceptor defects. The resistivity of Mo-doped and Ga- and Mo-codoped r-GeO$_2$ is very high at RT, but it decreases by 2–3 orders of magnitude upon heating to 300 °C, which corresponds to a thermal activation energy of ~0.17–0.18 eV. The results strongly motivate the research into the electrical properties and hole carrier activation energy of purely Ga-doped r-GeO$_2$ single crystals; when the incorporation of unintentional impurities can be avoided by choosing a more appropriate flux or alternative crystal growth method.

## Supporting Information

HR-XRD reciprocal space map analysis of a r-GeO$_2$ single crystal.

## Acknowledgements

We acknowledge the fruitful discussion with Prof. Takeshi Inoshita on the computational section of this work. The work at Tokyo Institute of Technology was supported by the MEXT Element Strategy Initiative to Form Core Research Center (JPMXP0112101001). C.A.N. acknowledges the support through a fellowship granted by the German Research Foundation (DFG) for proposal NI1834. T. Katase was supported by PRESTO, Japan Science and Technology Agency (JPMJPR16R1) and grant-in-aid for Scientific Scientists B (19H02425) from JSPS.

# Supporting Information:

# Shallow Valence Band of Rutile GeO$_2$ and P-type Doping

Christian A. Niedermeier,[1,*] Keisuke Ide,[1] Takayoshi Katase,[1] Hideo Hosono,[1,2] and Toshio Kamiya[1,2]

*Corresponding author: c-niedermeier@mces.titech.ac.jp

[1]Laboratory for Materials and Structures, Tokyo Institute of Technology, Yokohama 226-8503, Japan

[2]Materials Research Center for Element Strategy, Tokyo Institute of Technology, Yokohama 226-8503, Japan

Number of pages: 2

Table of contents.

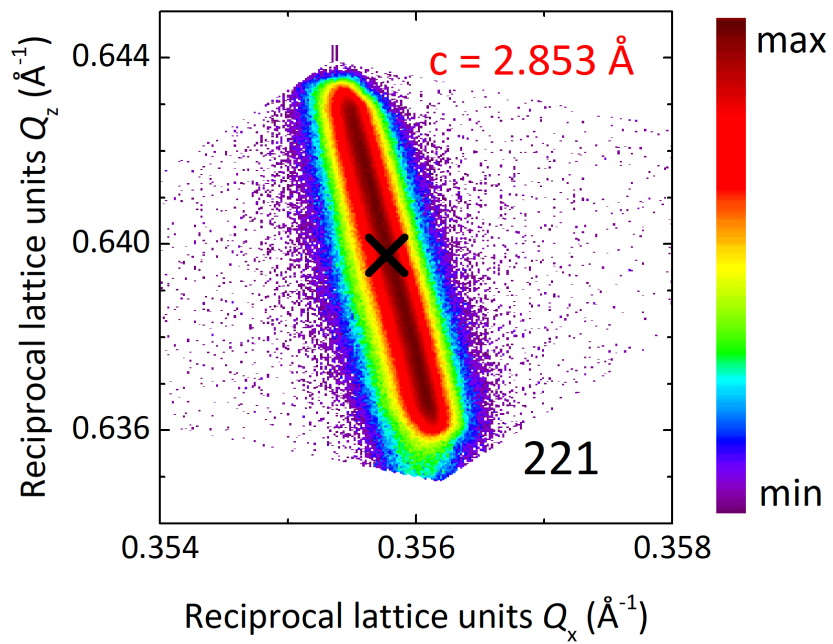

Fig. S1: Reciprocal space map of the in-plane 221 diffraction spot of a rutile $GeO_2$ single crystal with the surface normal aligned to the 110 crystallographic direction. The intensity is given on a logarithmic scale.